\documentclass[5p,twocolumn,times]{elsarticle}

\usepackage{lipsum}
\usepackage{lineno,hyperref}
\modulolinenumbers[5]
\usepackage[pdftex]{color}
\usepackage[font=footnotesize,labelfont=bf]{caption}
\usepackage[font=footnotesize,labelfont=bf]{subcaption}
\journal{Journal of \LaTeX\ Templates}

\usepackage{amssymb}

\biboptions{compress}

\usepackage[figuresright]{rotating}

\begin{document}

\begin{frontmatter}

\title{Spatial organisation plasticity reduces disease infection risk in rock-paper-scissors models}

\address[1]{Institute for Biodiversity and Ecosystem
Dynamics, University of Amsterdam, Science Park 904, 1098 XH
Amsterdam, The Netherlands}

\address[2]{School of Science and Technology, Federal University of Rio Grande do Norte\\
59072-970, P.O. Box 1524, Natal, RN, Brazil}

\author[1,2]{J. Menezes} 
\author[2]{S. Batista}  
\author[2]{E. Rangel}

\begin{abstract}
We study a three-species cyclic game system where organisms face a contagious disease whose virulence may change by a pathogen mutation. As a responsive defence strategy, organisms' mobility is restricted to reduce disease dissemination in the system. The impact of the collective self-preservation strategy on the disease infection risk is investigated by performing stochastic simulations of the spatial version of the rock-paper-scissors game.
Our outcomes show that the mobility control strategy induces plasticity in the spatial patterns with groups of organisms of the same species inhabiting spatial domains whose characteristic length scales depend on the level of dispersal restrictions. The spatial organisation plasticity allows the ecosystems to adapt to minimise the individuals' disease contamination risk if an eventual pathogen alters the disease virulence.
We discover that if a pathogen mutation makes the disease more transmissible or less lethal, the organisms benefit more if the mobility is not strongly restricted, thus forming large spatial domains.
Conversely, the benefits of protecting against a pathogen causing a less contagious or deadlier disease are maximised if the average size of groups of individuals of the same species is significantly limited, reducing the dimensions of groups of organisms significantly. Our findings may help biologists understand the effects of dispersal control as a conservation strategy in ecosystems affected by epidemic outbreaks.
\end{abstract}

\end{frontmatter}

\section{Introduction}
\label{sec1}
There is plenty of evidence that mobility plays a central role in species persistence, determining pattern formation and ecosystem stability \cite{ecology,Causes,MovementProfitable,Nature-bio}. Spatial interactions have been proved responsible for competing species' coexistence, as in the case of \textit{Escherichia coli} \cite{bacteria}. Experiments have shown that 
three strains of bacteria, whose dominance is nonhierarchical, survive because of a cyclic dominance, described by the rock-paper-scissors game rules. However, it has been proved that cyclic interactions are not sufficient to maintain biodiversity. This means that selection interactions must occur locally, leading to the formation of departed spatial domains \cite{Coli,Allelopathy}. Other biological systems, like Californian coral reef invertebrates and 
lizards in the inner Coast Range of California, have also provided 
evidence of the role of space in the maintenance of biodiversity in cyclic game systems \cite{lizards,coral}. In these systems, low mobility individuals interact with neighbours, producing a long-term coexistence; for high mobility values, conversely, spatial organisms' distribution tends to be homogeneous, resulting in biodiversity loss. 

Furthermore, behavioural ecology has demonstrated that mobility
plays an important role in the species adaptation to environmental changes \cite{foraging,butterfly,BUCHHOLZ2007401}. For example, the adaptive movement has been observed as either a survival strategy in dangerous situations or in foraging for search for prey or locations where the probability of species perpetuation is more propitious \cite{adaptive1,adaptive2,Dispersal,BENHAMOU1989375,coping}. Many organisms can scan the environment and interpret the signals captured from the neighbourhood to adjust their movement. The understanding of behavioural mobility strategies has helped the generation of sophisticated tools used by engineers to improve robots that imitate the animal behaviour \cite{animats}.

Scientists have shown that behavioural strategies 
can minimise the individual risk of being contaminated by a viral disease transmitted person-to-person \cite{social1,disease4,disease3,disease2,tanimoto}. For example, social distancing rules have been implemented worldwide, with individual and collective gains \cite{socialdist,soc,doi:10.1126/science.abc8881}. 
Furthermore, mobility restrictions have been shown efficient in decreasing the number of infected organisms and, consequently, minimising the social impact of epidemics on communities \cite{mr0,mr1,mr2,10.1371/journal.pone.0254403}. However, to guarantee the maximum efficiency of these measures in protecting against disease contamination, they are subject to adjustments if mutation alters the predominant pathogen causing the epidemics \cite{CAPAROGLU2021111246}.
The spatial organisation plasticity resulting from controlling the organisms' dispersal may be fundamental to the improve the efficiency of mitigation strategies against the disease with changing transmissibility and mortality \cite{plasticity2,Gene,mutating1,mutate2,plasticity1}.

In this work, we investigate a three-species cyclic game, where organisms cope with an epidemic of a deadly disease transmitted by neighbour agents. 
We aim to address the following questions: i) how do organisms' mobility restrictions control the spatial pattern plasticity by changing the size of departed spatial domains?; ii) what results does spatial pattern plasticity provide to prevent organisms' disease contamination?; iii) how do changes in the disease virulence (transmission and mortality rates) interfere with the dispersal reduction strategy?
 iv) how tight should organisms' mobility be restricted
to maximise that protection against infection of a varying virulence disease?

Our stochastic simulations are based on the May-Leonard implementation of the rock-paper-scissors Model, where organisms interact locally, but the number of individuals is not conserved \cite{tanimoto2,mobilia2,Szolnoki-JRSI-11-0735, Moura, Anti1,anti2,MENEZES2022101606,PhysRevE.97.032415,Avelino-PRE-86-036112,Bazeia_2017,PhysRevE.99.052310}. We assume a person-to-person disease transmission, which may affect all healthy organisms, irrespective of the species - no individual or species is immune to the viral infection \cite{combination,adaptive,rps-epidemy,epidemic-graphs,germen}. Once contaminated, individuals work as viral vectors, passing the disease to immediate neighbours before dying or being cured of the illness; cured individuals may be reinfected anytime.

The outline of this paper is as follows: the model is introduced in 
Sec.~\ref{sec2}, where the simulations are explained, and parameters are defined. We study spatial pattern plasticity and quantify the spatial domain scales in Sec.~\ref{sec3}. 
The interference of variations in the disease virulence in the population dynamics is investigated in Sec.~\ref{sec4}. In Sec.~\ref{sec5}, we study the adaptation in the mobility restrictions to respond to viral mutations changing the disease virulence. Finally, we discuss the results and present our conclusions in Sec.~\ref{sec6}.
\begin{figure}
\centering
\includegraphics[width=50mm]{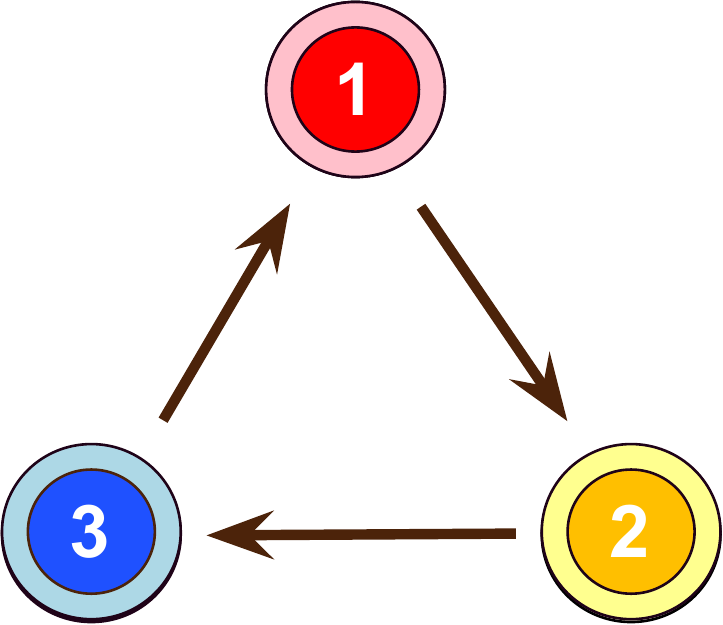}
\caption{Illustration of the spatial rock-paper-scissors model. Red, yellow, and blue circles represent organisms of species $1$, $2$, and $3$, respectively; dark and light colours stand for healthy and sick individuals, respectively.
The arrows show the cyclic selection dominance.}
	\label{fig1}
\end{figure}

\begin{figure*}
 \centering
        \begin{subfigure}{.29\textwidth}
        \centering
        \includegraphics[width=50mm]{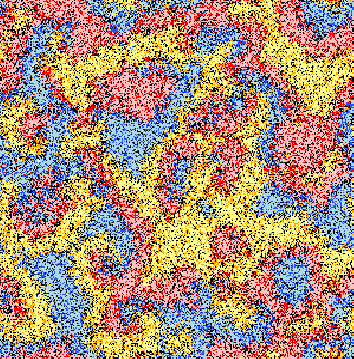}
        \caption{}\label{fig2a}
    \end{subfigure}
   \begin{subfigure}{.29\textwidth}
        \centering
        \includegraphics[width=50mm]{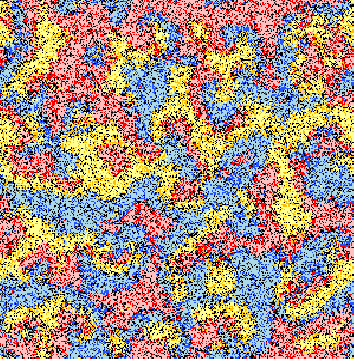}
        \caption{}\label{fig2b}
    \end{subfigure} 
           \begin{subfigure}{.29\textwidth}
        \centering
        \includegraphics[width=50mm]{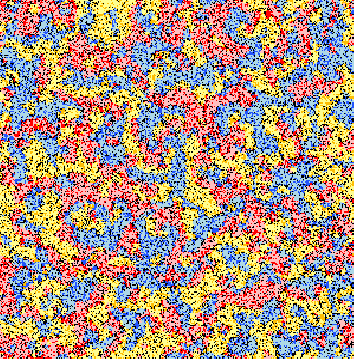}
        \caption{}\label{fig2c}
    \end{subfigure} 
\caption{Typical spatial patterns of the rock-paper-scissors model with a disease spreading for various slowness factors. The snapshots were captured from lattices with $300^2$ grid points, running until $t=5000$.  Figures a, b, and c show the snapshot at the end of the simulation for $\nu=0.0$, $\nu=0.50$, and $\nu=1.0$, respectively. Each organism is depicted with the colours in the scheme in Fig.~\ref{fig1}, with dark and light colours indicating healthy and sick individuals, respectively.
Black dots show empty spaces.}
  \label{fig2}
\end{figure*}

\begin{figure}[t]
\centering
\includegraphics[width=87mm]{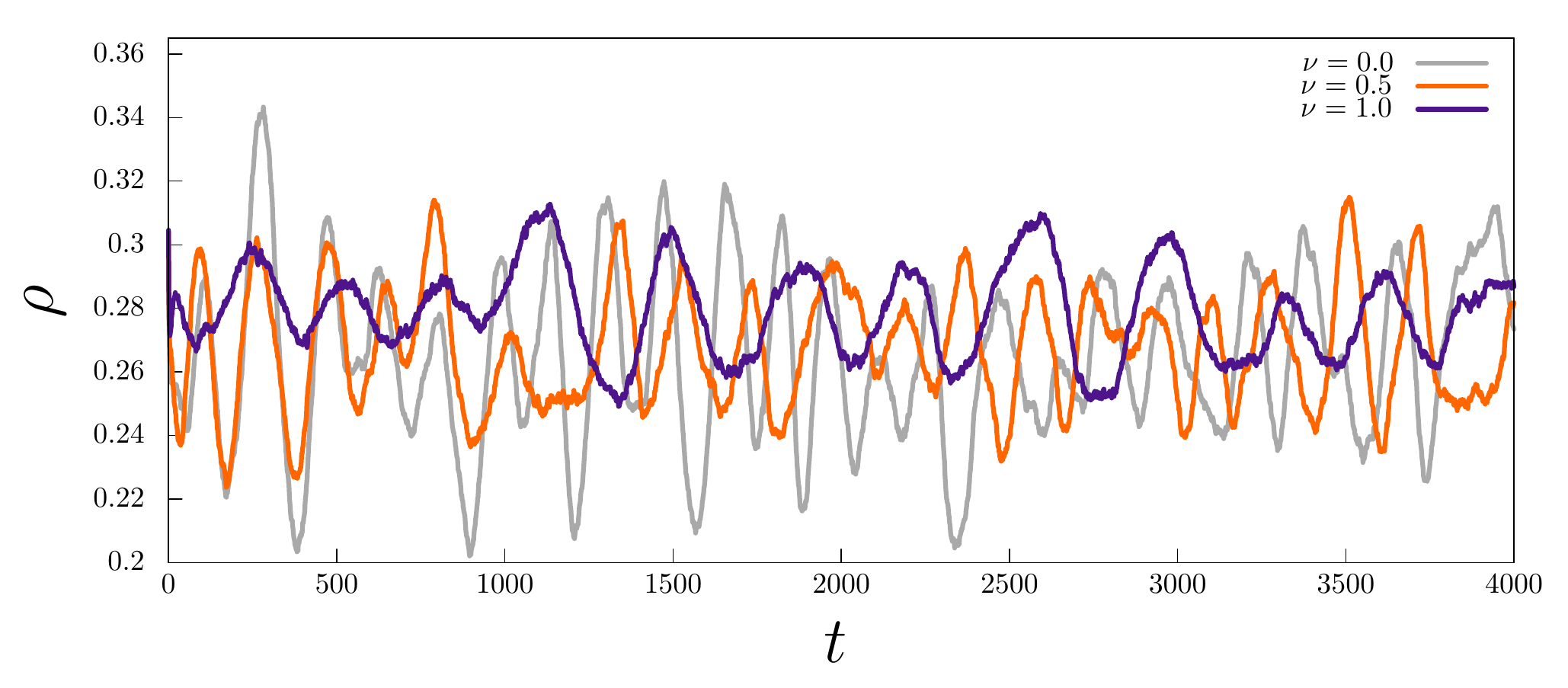}
\caption{The temporal dynamics of the fraction of the grid occupied by organisms of species $1$ during the simulations shown in 
Fig.~\ref{fig2}. Grey, orange, and purple lines depict the cases $\nu=0.0$, $\nu=0.50$, and $\nu=1.0$, respectively. }
	\label{fig3}
\end{figure}

\section{Our model}
\label{sec2}

We study a three-species cyclic model where selection dominance obeys the rock-paper-scissors model. In this popular game, scissors cut paper, paper wraps rock, and rock crushes scissors, as illustrated in Fig.~\ref{fig1}. Accordingly, red, yellow, and blue identify organisms of species $1$, $2$, and $3$, respectively; dark and light colours stand for healthy and sick individuals, respectively. The arrows indicate the cyclic species dominance, with organisms of species $i$ eliminating individuals of species $i+1$, with $i=i\,+\,3\,\alpha$, where $\alpha$ is an integer. 

We consider that a contagious disease, transmissible person-to-person, spreads through the system, affecting organisms of every species. 
We aim to understand how organisms' mobility can transform the spatial organisation to reduce the risk of disease contamination. 
We implement simulations where spatial pattern plasticity is controlled by restricting the organisms' mobility according to the disease virulence.

Our numerical implementation follows the May-Leonard model, where the total number of organisms is not conserved \cite{leonard}.
The simulations were performed in square lattices with periodic boundary conditions: individuals interact on a torus surface with $\mathcal{N}$ points. Each grid site contains at most one individual - the maximum number of individuals is $\mathcal{N}$. 
This 

The initial conditions are random: we allocate an organism of an aleatory species at each grid point. The initial number of individuals is the same 
for every species: $I_i \approx \mathcal{N}/3$, with $i=1,2,3$. Initially, the proportion of sick individuals is $1\%$, which is valid for every species. 

To describe the stochastic interactions implemented in our simulations, we first define the notation $h_i$ and $s_i$ to identify healthy and sick individuals of species $i$; the labelling $i$ stands for all individuals, irrespective of illness or health. The organisms' spatial distribution is altered by the implementation of one of the following interactions: 
\begin{itemize}
\item 
Selection: $ i\ j \to i\ \otimes\,$, with $ j = i+1$, where $\otimes$ means an empty space. A selection interaction results in an empty space in the grid site previously occupied by the individual of species $i+1$.
\item
Reproduction: $ i\ \otimes \to i\ i\,$. An offspring of species $i$ is produced to occupy the empty space.
\item 
Mobility: $ i\ \odot \to \odot\ i\,$, where $\odot$ means either an individual of any species or an empty site. Mobility happens when an individual of species $i$ switches grid site with another organism of any species or with an empty space.
\item 
Infection: $ s_i\ h_j \to s_i\ s_j\,$, with $i,j=1,2,3$. An ill individual of species $i$ transmits the virus to a healthy individual, irrespective of the species.
\item 
Cure: $ s_i \to h_i\,$. An ill organism of species $i$ is cured of the disease; once cured, the organism is vulnerable to being reinfected.
\item 
Death: $ s_i \to \otimes\,$. A sick individual of species $i$ dies because of complications of the disease; its position becomes an empty space.
\end{itemize}

In our stochastic simulations, the occurrence of a given interaction
depends on the set of real parameters: $S$ (selection rate), $R$ (reproduction rate), $M$ (mobility rate), $\kappa$ (infection rate), $\mu$ (mortality rate), and $\omega$ (cure rate). 
In addition, we implement the mobility restriction strategy by defining the slowness factor $\nu$, a real parameter that defines the percentual reduction in the mobility rate.
This means that the effective mobility rate is given by $(1-\nu)\,m$. 

The interactions were implemented by assuming the Moore neighbourhood; thus, an organism may interact with one of its eight immediate neighbours. The algorithm follows three steps: i) randomly choosing an active individual; ii) raffling one interaction to be executed; iii) drawing one of the eight nearest neighbours to suffer the interaction.
One time step is counted either if one interaction is implemented or if an organism is chosen to move but stays in its grid site because of the mobility restrictions. Our time unit is defined as the necessary time for $\mathcal{N}$ timesteps to occur. 
	
The density of organisms of species $i$ at time $t$ is defined as
\begin{equation} 
\rho_i (t)\,=\, \frac{I_i (t)}{\mathcal{N}},
\end{equation} 
with $i=1,2,3$, where $I_i(t)$ stands to the total number of individuals of species $i$ present in the lattice at time $t$. In addition, $\rho_0$ represents the density of empty spaces.
Because of the cyclic symmetry, the average value of the density of species is the same for all species; thus, we choose species $1$ to calculate the density of species $\rho$. 
We also define the densities $\rho_h (t)=I_h (t)/\mathcal{N}$ and $\rho_s (t)=I_s (t)/\mathcal{N}$
as the fraction of the grid occupied by healthy and sick individuals of species $1$, respectively.

\section{Spatial pattern plasticity}
\label{sec3}

\begin{figure}[t]
\centering
\includegraphics[width=87mm]{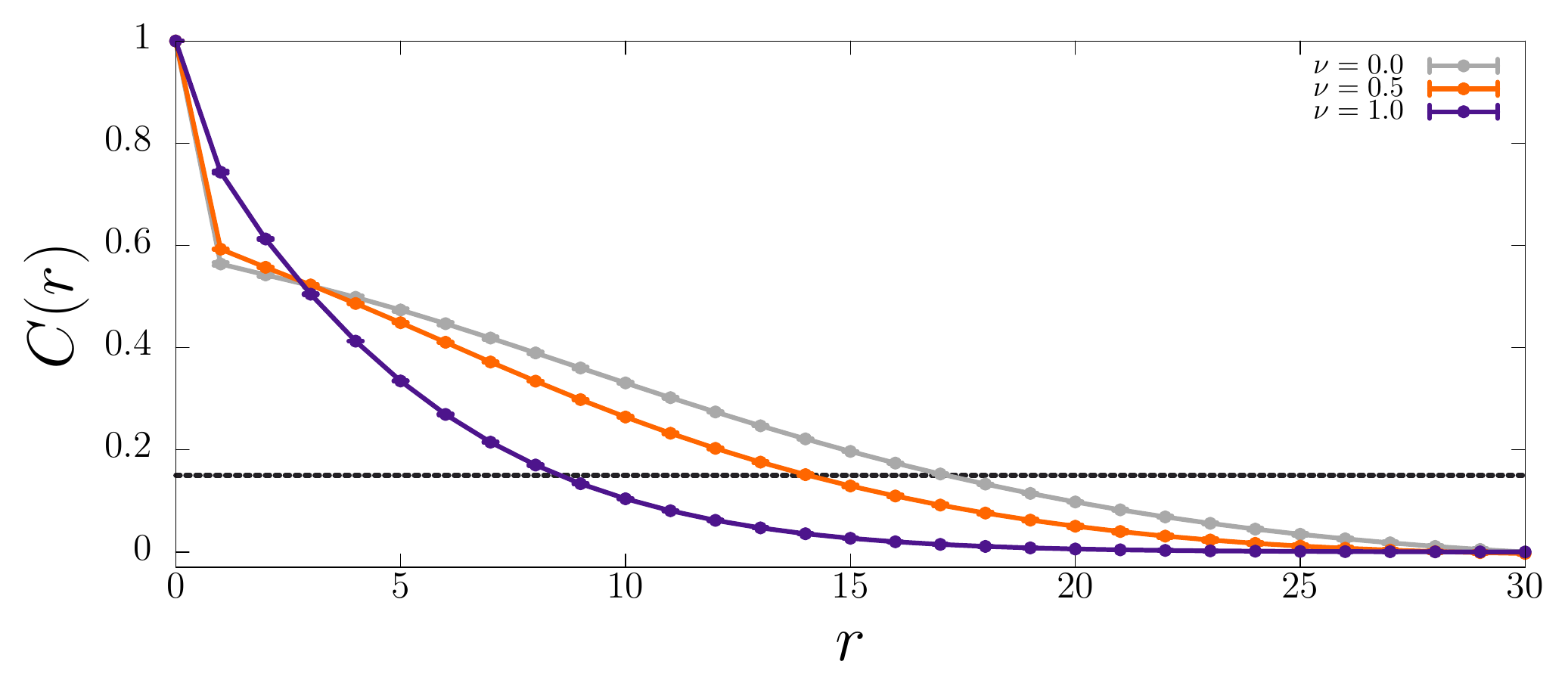}
\caption{Spatial autocorrelation function for various levels of organisms' mobility restrictions.
Grey, orange, and purple lines depict the cases $\nu=0.0$, $\nu=0.50$, and $\nu=1.0$, respectively. The error bars indicate the standard deviation in sets of $100$ simulations.}
	\label{fig3b}
\end{figure}

To understand how the organisms' mobility restriction strategy controls the spatial organisation plasticity during an epidemic, we first observe snapshots obtained from simulations for $\nu=0.0$, $\nu=0.50$, and $\nu=1.0$. The simulations were performed in lattices with $300^2$ grid sites until $t=5000$; the final spatial distributions were captured and shown in Figs. \ref{fig2a} ($\nu=0.0$), \ref{fig2b} ($\nu=0.50$), and \ref{fig2c} ($\nu=1.0$), respectively.
The colours follow the scheme in Fig.~\ref{fig1}: dark red, yellow, and blue dots show healthy individuals of species $1$, $2$, and $3$, respectively - light colours represent ill organisms of the respective species. The simulations ran for the set of parameters:
$S =R=1.0$, $M=\kappa=2.0$, and $C=\mu=0.2$.
The temporal change of the density of species during the entire simulations is depicted in Fig.~\ref{fig3}. Grey, orange, and purple lines indicate the fraction of the grid occupied by species $1$ in the simulations for $\nu=0.0$, $\nu=0.50$, and $\nu=1.0$, respectively.

After an initial transient pattern formation stage, healthy and sick individuals of the same species segregate in departed spatial domains. The cyclic selection dominance inherent in the rock-paper-scissors game leads to the arising of spiral waves, with individuals of the species occupying departed spatial domains forming the spiral arms.
The outcomes show that smaller groups of organisms 
as $\nu$ grows, implying that the average area of the single-species spatial domains decreases as mobility reduction accentuates - from Fig. \ref{fig2a} ($\nu=0.0$) to Fig. \ref{fig2c} ($\nu=1.0$). This is in agreement with the random walks theory; accordingly, the average area explored by individuals is proportional to the mobility rate \cite{mobilia2,random}. 
Fig.~\ref{fig3}, mobility restrictions slow the dynamics of the species populations: as $\nu$ grows, the amplitude and frequency of the species densities decrease because organisms are less exposed to being infected and eliminated by an enemy in the cyclic game; thus, the individuals' death risk drops.

\subsection{Typical spatial domain's characteristic length}
The snapshots in Fig.~\ref{fig2} show that the spatial organisation adapts to the organisms' velocity decrease. 
We aim to compute the changes in the scale of spatial domains occupied by each species resulting from the mobility restrictions. In this sense, we first compute the spatial autocorrelation function $C_i(r)$, with $i=1,2,3$, in terms of radial coordinate $r$ by introducing the function $\phi_i(\vec{r})$ that identifies the position $\vec{r}$ in the lattice occupied by individuals of species $i$ (healthy and ill individuals). Employing the Fourier transform
\begin{equation}
\varphi_i(\vec{K}) = \mathcal{F}\,\{\phi_i(\vec{r})-\langle\phi_i\rangle\},
\end{equation}
where $\langle\phi_i\rangle$ is the mean value of $\phi_i(\vec{r})$, we find the spectral densities
\begin{equation}
S_i(\vec{K}) = \sum_{K_x, K_y}\,\varphi_i(\vec{K}).
\end{equation}

Therefore,
\begin{equation}
C_i(\vec{r}') = \frac{\mathcal{F}^{-1}\{S_i(\vec{K})\}}{C(0)},
\end{equation}
can be rewritten as 
a function of the radial coordinate $r$:
\begin{equation}
C_i(r) = \sum_{|\vec{r}'|=x+y} \frac{C_i(\vec{r}')}{min\left[2N-(x+y+1), (x+y+1)\right]}.
\end{equation}
We calculate the mean autocorrelation function from a set of $100$ simulations with different initial conditions.
The realisations ran in lattices with $500^2$ grid sites, running until $t=5000$. We assumed the same set of parameters used in the results shown in Figs.~\ref{fig2} and \ref{fig3}.
Because of the symmetry of the rock-paper-scissors game, the typical spatial domains occupied by individuals of every species have the same average size, which is valid if the interaction rates are the same for every species. Therefore, we arbitrarily choose to compute the autocorrelation function for species $1$, defined as $C(r)$. We then use the threshold $C(l)=0.15$ to determine the characteristic length scale of the spatial domains, $l$.

First, we computed the spatial autocorrelation function $C(r)$, whose averaged results are depicted in Fig.~\ref{fig3b}; the error bars indicate the standard deviation.
Grey, orange, and purple lines show the autocorrelation function for the cases shown in 
Figs. \ref{fig2a} ($\nu=0.0$), \ref{fig2b} ($\nu=0.5$), and \ref{fig2c} ($\nu=1.0$), respectively. The dashed black line represents the threshold to calculate the characteristic scale $l$. The outcomes reveal that 
as dispersal restrictions increases, the organisms of the same species are less spatially correlated. This results from the arising of smaller single-species spatial domains, as shown in Figs.~\ref{fig2a} to \ref{fig2c}.

\begin{figure}[t]
\centering
\includegraphics[width=87mm]{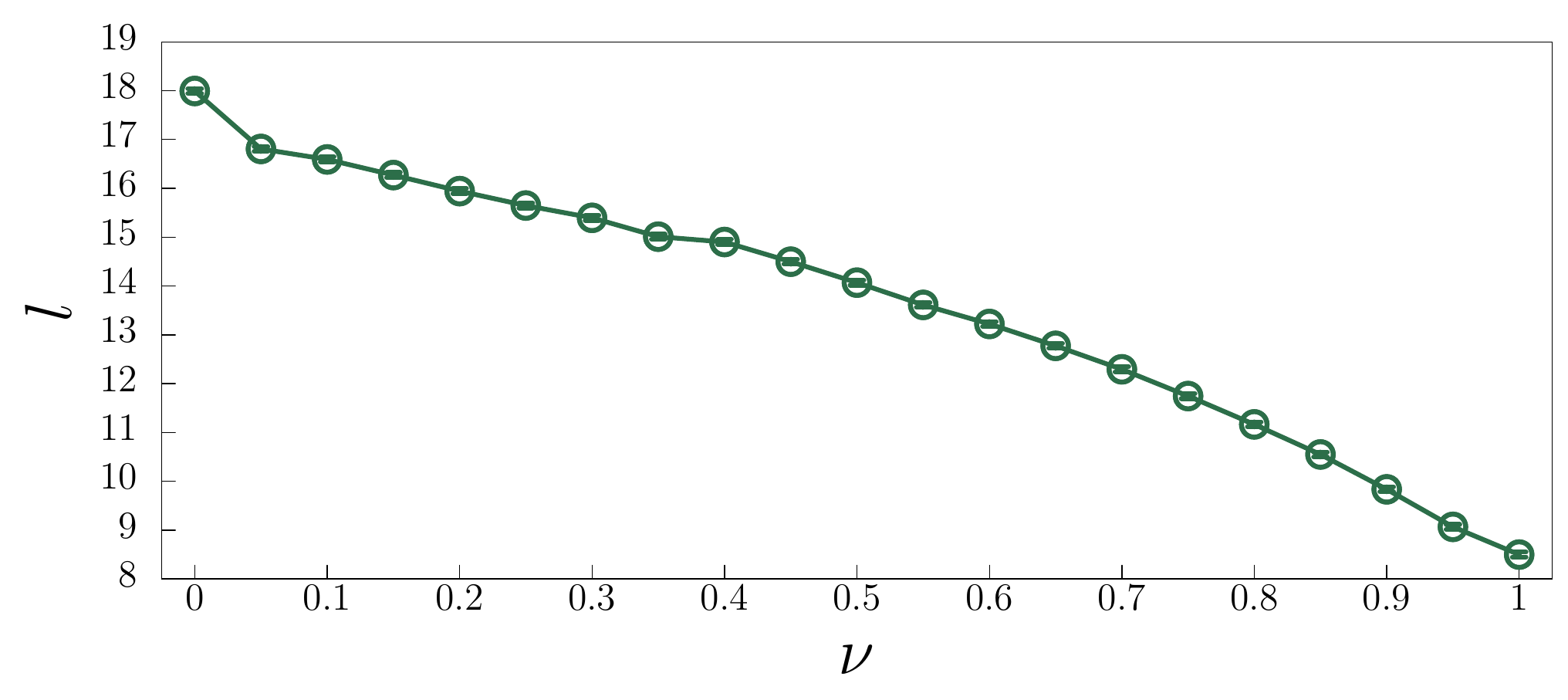}
\caption{Characteristic length scale of the typical single-species spatial domain as a function of the slowness factor. The outcomes were 
obtained by averaging a set of
$100$ simulations running in lattices with $500^2$ grid sites;
the error bars show the standard deviation.}
	\label{fig4}
\end{figure}

Using the threshold illustrated by the dashed black line in Fig.\ref{fig3b}, we 
quantified the spatial organisation plasticity induced by the individuals' mobility limitation in terms of the mean characteristic length scale $l$.
Figure~\ref{fig4} shows the mean value of $l$ for $\nu=0.0$ to $\nu=1.0$, in intervals of $\delta \nu =0.05$; the error bars show the standard deviation. 
The results show that $l$ nonlinearly decreases with $\nu$. In the limit case, where organisms are imposed to be static ($\nu=1.0$), the characteristic length scale of the typical single-species domains decreases by approximately $42\%$.

\section{Impact of disease virulence on population dynamics}
\label{sec4}

We now investigate the effects of variations in disease virulence caused by a virus mutation on population dynamics. For this purpose, we consider the following characteristics of the contagious disease:
\begin{itemize}
\item
Disease transmission: we performed a series of simulations for a wide range of infection rate, $\kappa$. If $\kappa$ grows, the virus mutation leads to a faster disease spreading.
\item
Disease mortality: we ran many simulations considering a wide interval of mortality rate, $\mu$. As $\mu$ grows, the mutation generates a deadlier disease.
\end{itemize}
\begin{figure}
 \centering
       \begin{subfigure}{.48\textwidth}
        \centering
        \includegraphics[width=85mm]{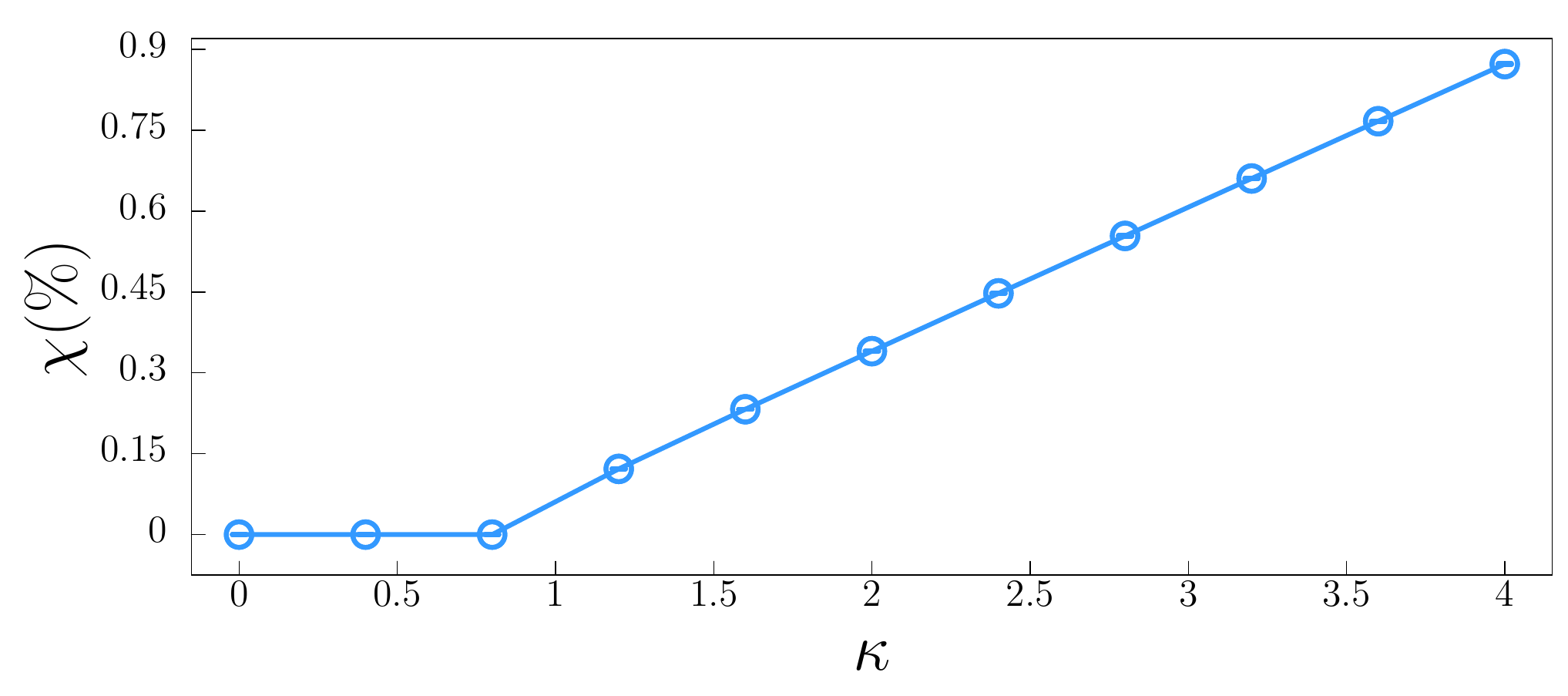}
        \caption{}\label{fig5a}
    \end{subfigure}\\
           \begin{subfigure}{.48\textwidth}
        \centering
        \includegraphics[width=85mm]{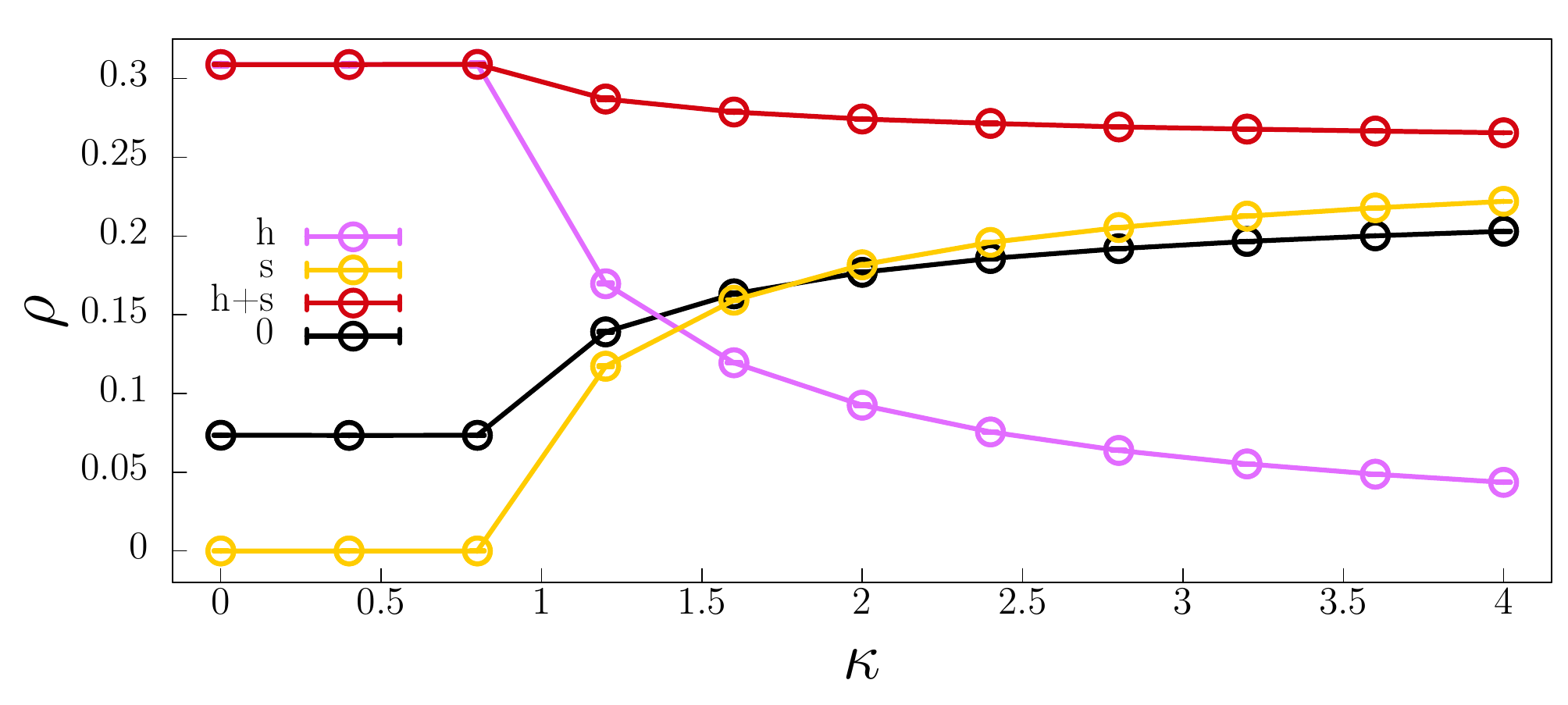}
        \caption{}\label{fig5b}
    \end{subfigure}
\caption{Relative change in organisms' infection risk and species densities as a function of the infection rate.
Figure \ref{fig5a} shows how $\chi$ varies with $\kappa$ while Fig.~\ref{fig5b} depict the consequences on the specie density (red line), and the fractions of healthy (light purple line), sick (yellow line) organisms; the density of empty spaces is depicted by the black line.
The outcomes were averaged from a set of $100$ simulations, running in lattices with $500^2$ grid sites - the error bars indicate the standard deviation.}
  \label{fig5}
\end{figure}

\begin{figure}[t]
 \centering
       \begin{subfigure}{.48\textwidth}
        \centering
        \includegraphics[width=85mm]{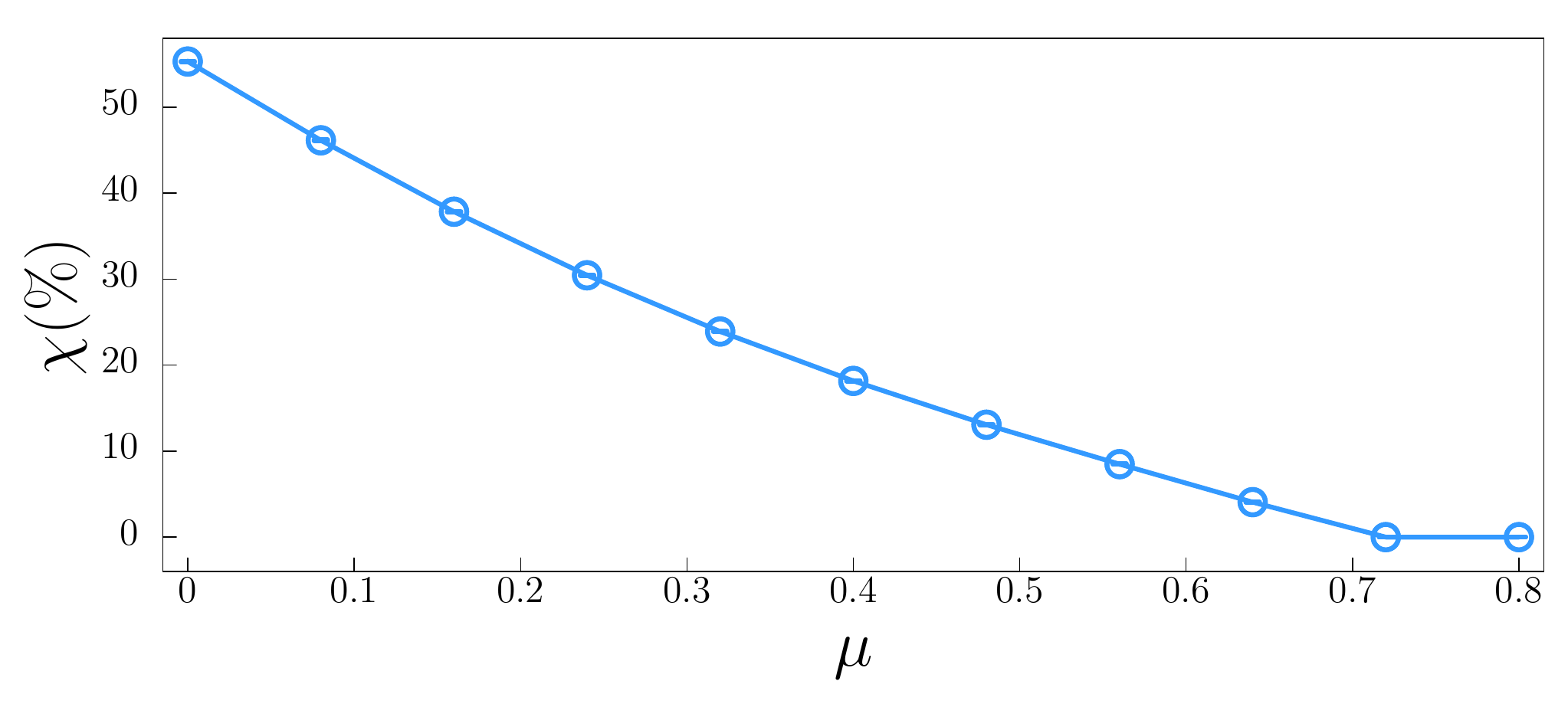}
        \caption{}\label{fig6a}
    \end{subfigure}\\
           \begin{subfigure}{.48\textwidth}
        \centering
        \includegraphics[width=85mm]{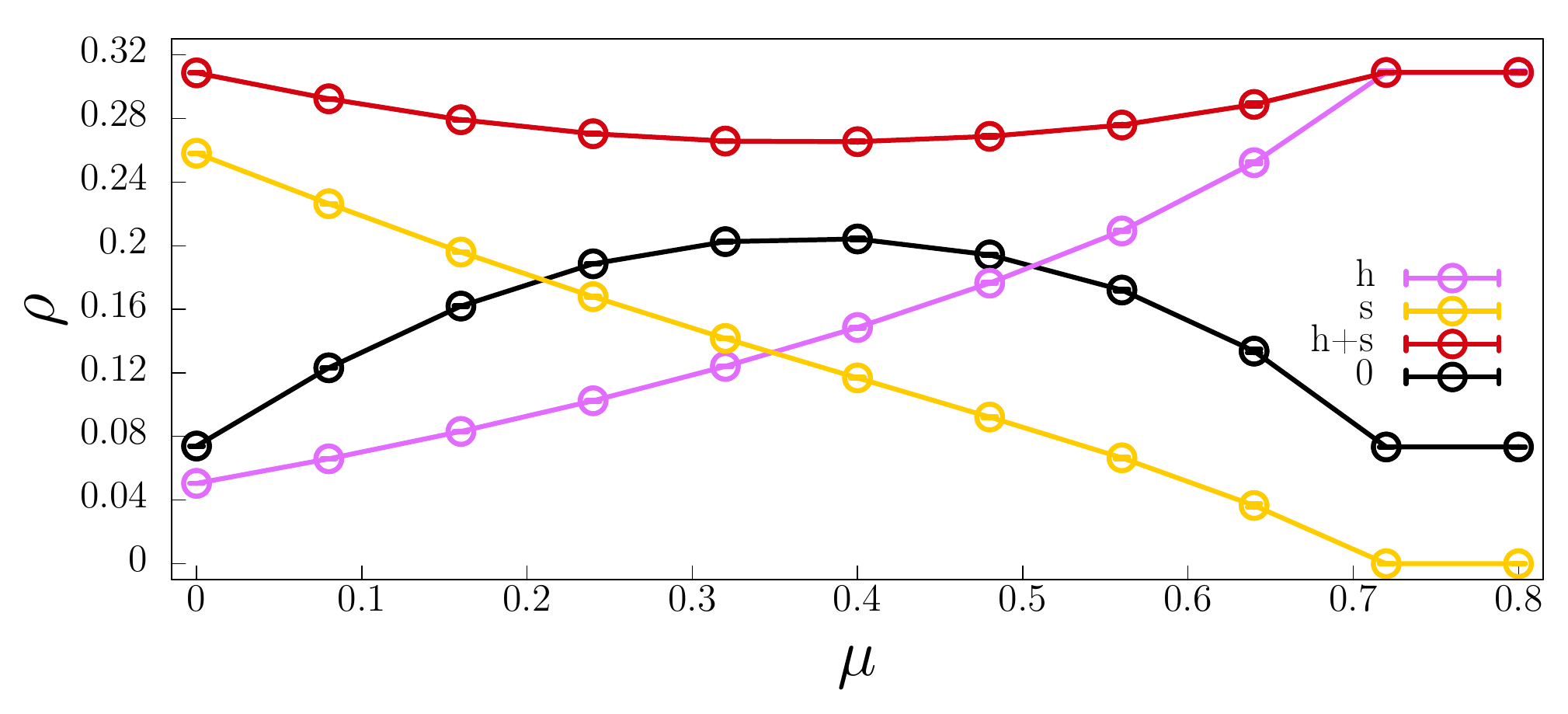}
        \caption{}\label{fig6b}
    \end{subfigure}
\caption{Relative change in organisms' infection risk and species densities as a function of the mortality rate.
Figure \ref{fig6a} shows how $\chi$ varies with $\mu$. 
Fig.~\ref{fig6b} depict the variation in the specie density (red line); the purple and yellow lines represent the densities of healthy and sick individuals, respectively, while the black line depicts the density of empty spaces.
The outcomes were averaged from a set of $100$ simulations, running in lattices with $500^2$ grid sites; the error bars show the standard deviation.}
  \label{fig6}
\end{figure}
\begin{figure}
 \centering
       \begin{subfigure}{.48\textwidth}
        \centering
        \includegraphics[width=85mm]{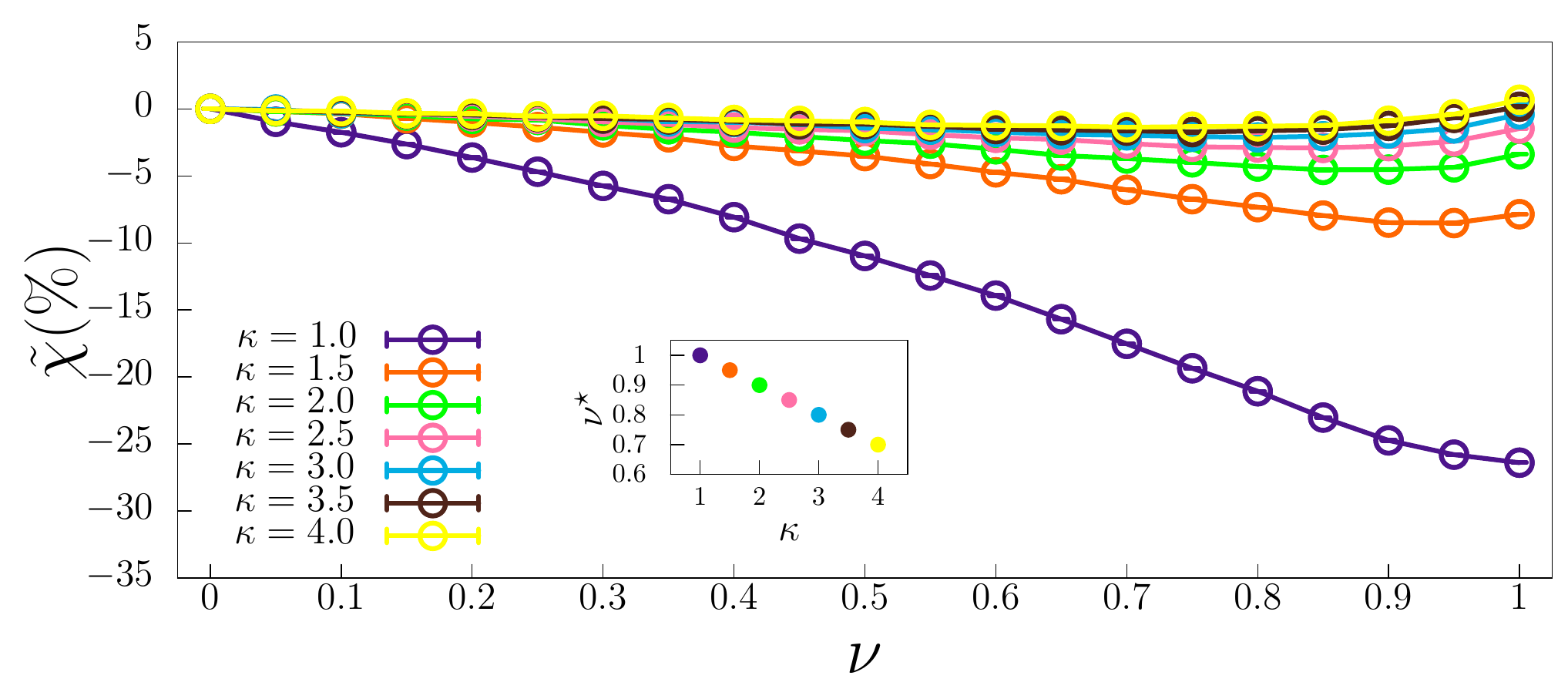}
        \caption{}\label{fig7a}
    \end{subfigure}
           \begin{subfigure}{.48\textwidth}
        \centering
       \includegraphics[width=85mm]{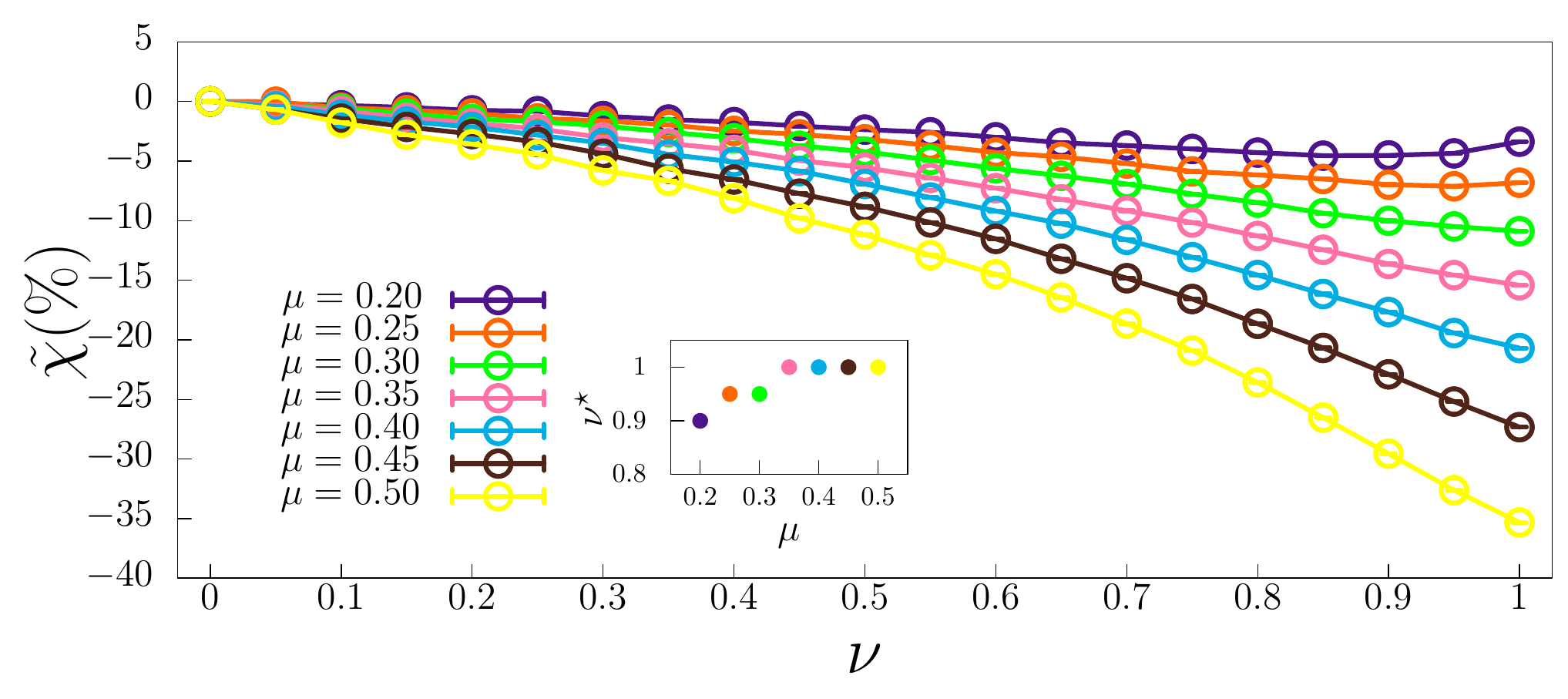}
        \caption{}\label{fig7b}
    \end{subfigure}
\caption{Relative change in organisms' infection risk 
as a function of the slowness factor for a variety of disease virulence. Figures \ref{fig7a} and \ref{fig7b} shows the dependence of $\tilde{\chi}$ in the infection and mortality rates, respectively.
The results were averaged from a set of $100$ realisations, running in lattices with $500^2$ grid sites; the error bars indicate the standard deviation. The inset figures show the optimum slowness factor $\nu^{\star}$, with the colours representing the respective $\kappa$ and $\mu$.}
  \label{fig7}
\end{figure}
 
We introduce the individuals' infection risk
$\chi_i(t)$, as the probability of a healthy organism of species $i$ being contaminated at time $t$.
The infection risk is implemented following the following steps:
i) the total number of healthy individuals of species $i$ when each unit time begins is saved; ii) the number of individuals of species $i$ infected during the unit time is computed; 
iii) the infection risk, $\chi_i$, with $i=1,2,3$ is found by the ratio between the number of infected individuals and the initial number of healthy individuals.
Throughout this work, $\chi$ stands for the infection risk, valid for every species because of the cyclic model symmetry - the calculation is made using the data from species $1$.

We performed a series of $100$ simulations in grids with $5000^2$, running until $t=5000$. 
We avoid the high fluctuations in the species densities inherent to the pattern formation process (in the initial simulation stage) by computing the mean infection risk and species densities using the outcomes from the second simulation half. The simulations were performed for the same sets of parameters as in the previous section, except for $\kappa$ and $\mu$, which are specified in Figs.~\ref{fig5} and \ref{fig6}, respectively. The slowness factor is assumed to be $\nu=0.6$.

The mean infection risk $\chi$ is depicted in Figs.~\ref{fig5a} and ~\ref{fig6a} for a wide range of
infection and mortality probabilities, respectively; the error bars indicate the standard deviation. Additionally, we calculated the mean value of the densities of organisms of a single species and empty spaces. The outcomes are depicted by light purple (healthy individuals), yellow (sick individuals), red (healthy and sick individuals), and black lines (empty spaces) in Fig.~\ref{fig5b} and Fig.~\ref{fig6b} as functions of the infection, and mortality probabilities, respectively.

\subsection{Disease Transmission}
As all simulations started with only $1\%$ of organisms being sick, the minimum infection rate that causes the
disease dissemination throughout the lattice is 
$\kappa=0.8$. This means that for $\kappa<0.8$, individuals of species $i$ die only if eliminated by organisms of species $i-1$ in the spatial rock-paper-scissors game. Therefore, the organisms' death risk is minimum. In this case, the average value of the species density is maximum since all organisms are healthy (light purple line in Fig.~\ref{fig5b}).

For $\kappa\geq0.8$, disease transmission is sustained, with the individuals' infection risk growing approximately linearly in $\kappa$, as depicted in Fig.~\ref{fig5a}.
This yields the species densities to decline because of the increase in the proportion of sick individuals and the consequent growth in the density of empty spaces (yellow and black lines in Fig.~\ref{fig5b}). As $\kappa$ grows, the relative density of healthy individuals decreases; therefore, the disease outbreak affects fewer and fewer healthy individuals. 
 
\subsection{Disease Mortality}
The results reveal that if the disease does not provoke the death of sick individuals ($\mu=0$),
the proportion of ill organisms is maximum; namely, $\rho_s\, =\,0.836 \,\rho$. Because of this, 
the infection risk is maximum, as depicted in Fig.~\ref{fig6a}. 

As $\mu$ increases, the number of organisms dying because of the disease rises, reaching the maximum value for $\mu=0.4$. As depicted by the black line in Fig.~\ref{fig6b}, the density of empty spaces is maximum for $\mu=0.4$, with the density of species $\rho=\rho_h+\rho_s$ being minimal.
For $\mu>0.4$, the increasing proportion of viral vectors dying before transmitting becomes counterproductive for the disease spreading. Thus, the disease transmission weakens to higher mortality rates, being eradicated for $\mu=0.72$. The results in Fig. ~\ref{fig6b} also show that the decrease in density of ill organisms is approximately linear in $\mu$.
\section{Adaptation of the mobility restriction strategy}
\label{sec5}

Finally, we aim to quantify the benefits of the spatial organisation plasticity as a self-preservation strategy against disease contamination. We assume that a pathogen mutation modifies the disease virulence and triggers the organisms' mobility restrictions to adjust the size of the single-species spatial domains. The goal is to maximise the relative decrease in individuals' disease contamination risk. For this purpose, we define the optimum slowness factor $\nu^{\star}$ as the level of the mobility restrictions that minimises the infection risk for varying $\kappa$ and $\mu$.

We performed two experiments to vary the disease transmission and mortality and computed the relative change in the infection risk: $\tilde{\chi}=(\chi-\chi_0)/\chi_0$, where $\chi_0$ is the infection risk for $\nu=0$. 
For each set of parameters, we ran $100$ simulations in lattices with $500^2$ grid points, starting from different initial conditions and running until $t=5000$. Except for $\kappa$ and $\mu$, whose values are shown in Figs. \ref{fig7a} and \ref{fig7b}, all parameters are the same as in the previous sections.

Figure \ref{fig7a} shows how the relative infection reduction, $\tilde{\chi}$ depends on $\nu$ for seven infection rates: $\kappa= 1.0$ (purple line), $\kappa= 1.5$ (orange line), $\kappa=2.0$ (green line), $\kappa= 2.5$ (pink line), $\kappa= 3.0$ (blue line), $\kappa= 3.5$ (brown line), and $\kappa=4.0$ (yellow line). 
The optimum slowness factor $\nu^{\star}$ depicted in the inset figure represents the ideal mobility reduction for the respective $\kappa$.

We first observe that the efficiency of the dispersal reduction tactic is more relevant for low $\kappa$. Furthermore, as less transmissible 
the disease becomes, the slower individuals should move to maximise the gain in protection against infection. For $\kappa=1.0$, being static ($\nu^{\star}=1.0$) brings the most significant reduction in $\chi$ - as depicted by the purple dot in the inset figure.
In contrast, if the mutation makes 
the disease more transmissible, it is more advantageous if the dispersal restrictions are partially released. As indicated in the inset figure, for $\kappa=4.0$, the optimum slowness factor is $\nu^{\star}=0.7$.

Concerning the variation in the disease mortality, we computed the relative variation in infection risk for seven cases: $\nu= 0.20$ (purple line), $\nu= 0.25$ (orange line), $\nu= 0.30$ (green line), $\nu= 0.35$ (pink line), $\nu= 0.40$ (blue line), $\nu= 0.45$ (brown line), and $\nu=0.50$ (yellow line), as shown in Fig.~\ref{fig7b}. The inset figure shows $\nu^{\star}$ for each value of $\mu$.

For a low mortality disease, the results indicate that $\nu^{\star}=0.9$ is the mobility restriction that makes the spatial organisation plasticity to provide maximum protection against disease contamination. However, once the disease mortality rises, the profit in reduced infection risk is maximised if the dispersal limitations accentuate. According to our results,
for $\mu\geq 0.30$, the best results are achieved for $\nu^{\star}=1.0$.
\section{Discussion and Conclusions}
\label{sec6}
We run simulations of the spatial version of the rock-paper-scissors rules, where organisms face an infectious disease outbreak. Running stochastic simulations, we first investigated how the individuals' mobility restriction strategy induces plasticity in the individuals' spatial organisation.
For this purpose, we quantified how the characteristic length scale of the typical single-species spatial domains decreases when individuals are constrained to explore a smaller fraction of the grid per unit time \cite{mobilia2,random}. 

To understand how changes in the disease virulence caused by the pathogen mutation affect population dynamics, we simulate the 
spreading of less severe and more aggressive diseases. The outcomes show that: 
\begin{itemize}
\item
If the new virus provokes a more transmissible disease, the proportion of sick individuals increases; thus, the number of individuals transmitting the virus grows. However, the situation is asymptotically stabilised because of the many deaths of sick individuals that control the transmission rate.
\item
Infection risk drops as the mortality rate grows because more deaths of sick organisms result in fewer virus vectors. 
Furthermore, we found that the population decline caused by the increasing number of fatalities is reverted if the pathogen mutation generates a very high mortal disease.
In this case, the initial extremely high number of deaths brings a consequent drastic drop in the number of individuals transmitting the disease. 
\end{itemize}

Our findings show the benefits of the mobility restriction
is affected by alterations in disease virulence. Therefore, the individuals' dispersal limitation strategy must adapt if pathogen mutation changes disease transmission or mortality. 
This means that if organisms are sensitive to changes in disease virulence, adapting the mobility rhythm, the spatial organisation plasticity ensures a maximum reduction in the infection risk.

The results revealed that the relative reduction in the infection risk decreases significantly if the disease becomes less transmissible or deadlier. In this case, organisms gain more protection if movement restrictions are tighter, creating spatial domains with shorter characteristic length scales.
On the contrary, faced with a more transmissible or less lethal illness, individuals' dispersal restrictions must be released, allowing the formation of larger groups of individuals of the same species. 

Although, in our model, a sick organism becomes susceptible to being reinfected when recovered from the disease, all conclusions hold if a temporary immunity is given to cured individuals. In this case, the disease transmission chain is weakened since the virus finds obstacles to spread. This introduces a time delay in the transmission process, thus decreasing the average infection risk. In this scenario, mobility restrictions also work as an effective collective survival strategy, inducing spatial organisation plasticity that reduces the individuals' disease contamination risk.

Our outcomes can be generalised to cyclic game systems with an arbitrary odd number of species \cite{Avelino-PRE-86-036112}, where organisms adapt their mobility rate to escape being infected or eliminated by enemies \cite{TENORIO2022112430,Menezes_2022}.
Furthermore, besides being interpreted as an evolutionary behavioural strategy that organisms perform to protect themselves from disease surges, our results can be helpful to ecologists in creating ecosystem conservation strategies aiming to protect biodiversity from epidemic outbreaks \cite{combination,adaptive}.

\section*{Acknowledgments}
We thank CNPq, ECT, Fapern, and IBED for financial and technical support.
\bibliographystyle{elsarticle-num}
\bibliography{ref}

\end{document}